# Multitube monitors: a new-generation of neutron beam monitors

F. Lafont[a1,], D. Barkats[a], J-C. Buffet [a], S. Cuccaro[a], B. Guerard [a], C-C Lai[b], J. Marchal[a], J. Pentenero[a], N. Sartor[a], R. Hall-Wilton[b], K. Kanaki[b], L. Robinson[b] and P-O. Svensson[b]

[a] *Institut Laue Langevin,*
*71 Avenue des martyrs 38042 Grenoble, France*

[b] *European Spallation Source ERIC,*
*Box 176, SE-221 00, Lund, Sweden*
*E-mail*: lafont@ill.fr

ABSTRACT: With the renewal of many neutron science instruments and the commissioning of new neutron facilities, there is a rising demand for improved neutron beam monitoring systems with reduced beam perturbations and higher counting rate capability. Fission chambers are the most popular beam monitors; however, their use on some instruments may be prevented by the background generated by fast neutrons emitted during neutron captures in $^{235}$U and by neutrons scattered in the material of the fission chamber. Multitube detectors, on the other hand, offer a good alternative with minimum beam perturbations. The purpose of this paper is to report and analyse the results of the measurements performed with several Multitubes used for beam monitoring. We show that the transparency of Multitube beam monitors is 97.6 $\pm$ 0.4 %, and that their detection efficiency is uniform, with a deviation from the mean value < 0.7 %. A counting rate reduction of 10 % due to pile-up effects is measured at a rate of 550 kHz. In addition to neutron beam intensity monitoring, the Multitube can be configured for 1-dimensional or 2-dimensional localisation. We present the preliminary results of these additional functionalities.

---

[1] Corresponding author.

**Contents**

# 1. Introduction

Transmission beam monitors are used to correct for the neutron flux variations in the experimental data collected on neutron instruments. These variations in flux are due to the instrumental conditions or to the neutron source itself and thus, they are a key element for many neutron instruments. There are two types of transmission beam monitor used at the Institut Laue Langevin neutron research facility (ILL): fission chambers and gas proportional counters. Neutron captures take place in a $^{235}$U film in the first case, and in either $N_2$, $^3$He, $^{10}$BF$_3$ gas or $^{10}$B-based films in the latter. The large Q-value of the fission reaction in $^{235}$U allows the fission chamber to be used in counting mode without gas amplification; this ensures counting stability and low gamma sensitivity, even in harsh irradiation environments. However, tests performed on some commercially available fission-based monitors show that the absorption and scattering reactions in the monitor materials, in particular the entrance window, contribute significantly to a reduction of the neutron beam intensity [1]. The resulting background radiation noise places constraints on the radiation shielding of the instrument, and in addition the fast neutrons produced in the reaction with uranium are not easily stopped. In [2] it is shown that fission chambers commercially available have a non-uniform response both in terms of efficiency and transparency to neutrons. Single wire proportional counters based on gas converter have a better detection uniformity than fission chambers, but they are limited in counting rate due to the dead time of neutron events, which is typically 5 microseconds in average [3], about a factor of 10 more than with a fission chamber. They also produce a high background due to the interaction of neutrons in the windows.

The Multitube detectors presented in this paper have been designed specifically for neutron beam monitoring. The objective of this study is to measure the performance of this new type of monitor in terms of beam transmission, response uniformity, and counting stability, in particular at high neutron rate, and in a harsh gamma environment.

# 2. Description of the Multitube beam monitor

## 2.1 Mechanical design

A Multitube detector is made of several gas proportional counter tubes sharing the same volume of gas. The tubes can either be made in stainless steel and welded onto gas tight vessels, or they can be machined using wire spark erosion machining (SEM) in a block of aluminium. In the latter case, the detector is called a MAM (Monoblock Aluminium Multitube). Multitube detectors already in operation are either stainless-steel based (ILL instruments IN5, IN13, and PANTHER) or MAM based (ILL instruments Figaro, D17, D33, IN16B, and D3). LLB instruments PAXY and PA20 are also equipped with MAM detectors and more recently the instrument Platypus [4] at ANSTO.

The MAM technology provides several advantages for small- and medium-sized detectors compared to stainless-steel tubes: they are mechanically more precise and the tubes can have a smaller pitch, down to 3 mm. Their section can be rectangular, allowing for a minimum dead zone between them and, finally, neutron scattering is reduced thanks to the use of thin aluminium walls.

The Multitube beam monitors currently used at the ILL are MAMs with tubes of 90 mm to 220 mm sensitive length. The monitor studied in this paper contains 10 tubes of 7.5 mm × 7.5 mm section at a pitch of 7.9 mm, and an active length of 90 mm. As for the other MAMs produced at the ILL, their tubes are machined by wire spark erosion in a single block of aluminium 5083 (Figure 1). This configuration provides the double advantage of strong rigidity and high transparency to neutrons. As regards its materials budget, with its two 0.5 mm thick windows and the 0.4 mm thick internal walls a MAM monitor is equivalent to an aluminium plate of 1.38 mm thickness. The aluminium vessel is sealed with two flanges equipped with knife-edge seals ensuring excellent gas tightness.

The monitor was initially developed to provide a beam monitor compatible with operations in a TOF vacuum chamber. This type of design can sustain high differential pressures, with negligible deformation of the windows ($<10^{-2}$ mm for a differential pressure of 1 bar). The detector vessel can be pumped down without the need to reduce the outside pressure; this is particularly convenient for outgassing the vessel before filling with the detection gas. The maximum internal pressure applicable is 20 bars; such a high pressure is not useful for beam monitoring, but it can be useful for other applications apart from beam monitoring, if a high detection efficiency is required.

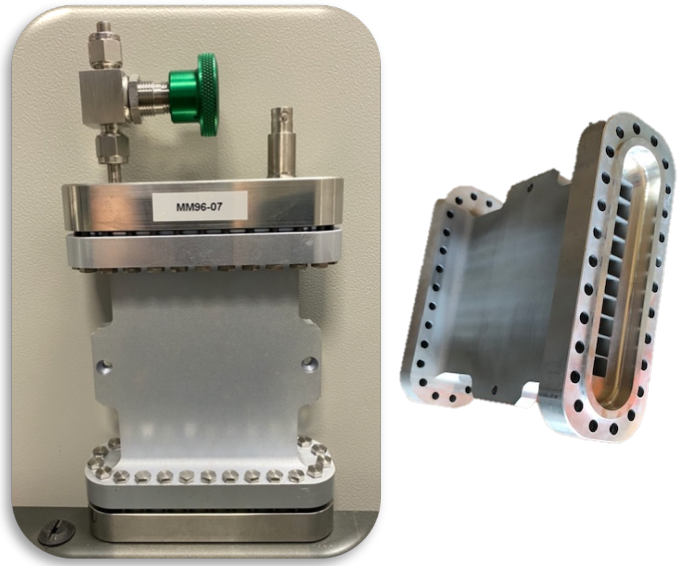

**Figure 1.** Picture of a 76x96 mm-active area Multitube monitor body with and without flanges (body)

The fact that the vessel of the Multitube is insensitive to gas pressure variations allows the anode wires to be positioned with high accuracy inside the tubes, resulting in a uniform amplification gain over the whole sensitive area.

**2.2 Materials**

The materials used for the MAM beam monitors have been chosen carefully, to ensure stability and compatibility with most experiments, including those using polarized neutrons. The main body of the monitor is made of 5083 Aluminium alloy to provide good transparency to neutrons and good mechanical performances. It is machined with high accuracy by SEM and the surface of the aluminium is chemically treated to prevent oxidation. The insulators are ceramic, while all the other parts, including the flanges, valves and feedthroughs are made of 316L stainless steel for low magnetic permeability. Only the valve bellows, in 321 stainless steel, are of slightly higher permeability. Titanium screws are used to seal the flanges. The anodes are 15-$\mu$m gold-plated tungsten-rhenium wires crimped with small copper tubes. Each monitor is outgassed for a week in an 80°C oven before being filled with the detection gas. A test of the magnetic impact of these devices was performed on the D7 instrument at the ILL with the beam monitor mounted close to the neutron polarizer; in this configuration, the beam monitor did not induce any noticeable depolarization of the neutrons.

## 2.3 Electronics read-out schemes

In standard operation mode, the anode wires are connected together on the same side with a conductive wire and read out through a single MHV connector; thus, the monitor acts as a counter (0-D). The signal is then amplified with homemade bipolar preamplifiers with a short integration time constant (110 ns or less) and can be read out with a Multi-Channel Analyzer (MCA) for example.

In order to measure the number of counts on each tube, a one-dimensional (1-D) localisation read-out system can be used, where neighbouring anode wires are connected together with 1-k$\Omega$ resistors (Figure 2, left). It is also possible to configure the monitor for two-dimensional (2-D) localisation by replacing the tungsten wires by resistive wires (e.g. Stablohm 710 wires of 20 μm diameter with a resistance of 38 $\Omega \cdot cm^{-1}$), and by connecting these serially via 51 $\Omega$ resistors in a zig-zag layout; this is equivalent to a single resistive line of 5.5 k$\Omega$ (Figure 2, right)

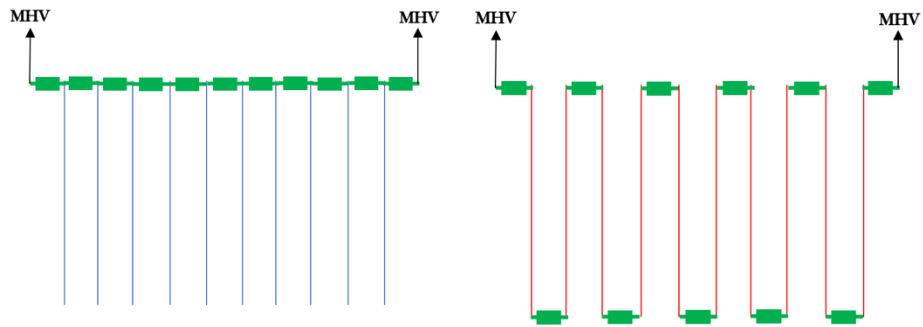

**Figure 2.** Principle schemes of the connections of the anode wires inside the Multitube monitor for 1-D (left) and 2-D (right) localization. Blue wires are conductive while red ones are resistive. Green boxes are the resistors.

In either case, 1-D and 2-D, the resistive chain is read out at both ends by using two MHV feedthroughs each one connected to a preamplifier, and the position of a neutron along the resistive line is processed by charge division electronics (CAEN V1740D). In 2-D localisation, a high amplification gain is required in order to measure the position along the tubes with adequate spatial resolution, whereas a moderate amplification gain is sufficient to determine the position of the fired tube in the 1-D localisation. For the measurement described in Sections 4 and 5 the electronics was specified for 1-D, but was used for both the 1-D and 2-D measurements.

## 2.4 Neutron converter and detection gas

Ar-$CO_2$ (90:10) is used as the stopping/quenching gas. Several gas mixtures have been tested to minimize dead-time. $^3$He or $N_2$ gas or boron-based thin films are used to convert neutrons into ionizing particles according to the following reactions:

$^3$He (n,p) $^3$H + 764 keV (5330 barn)   (1)
$^{14}$N (n,p) $^{14}$C + 626 keV (1,9 barn)   (2)
$^{10}$B (n,α) $^7$Li (+ 2.3 MeV + γ 93.6 %) (+ 2.78 MeV 6.4 %) (3840 barn)   (3)

The use of boron films is an attractive solution to facilitate maintenance when vacuum and high-purity gas handling systems are not accessible. MAM monitors have been tested with boron films of different thicknesses mounted inside each tube. This approach provides counter tubes of different efficiencies covering a broad range of beam intensities. When exposed to the neutron beam, some of these counters (the most efficient ones) might saturate, whereas some others (the less efficient ones) might produce data with poor statistics due to insufficient counting rate. Between these two extrema, there will be at least one counter with the adequate detection efficiency. The tubes cannot be gathered on a single read-out channel, otherwise the signal would be saturated by the most efficient counters. This type of multichannel universal beam monitor is promising, but it has the constraint that each counter tube has to be connected to an individual read-out channel.

## 3. Characterization

The higher the counting rate limit of the monitor, the larger the neutron flux range covered. A large flux range is particularly important for instruments with changing beam conditions; furthermore, the higher the neutron detection efficiency, the lower the contribution of gamma-rays to the total rate. As a result, counting rate capability is one of the key performance parameters of a neutron beam monitor. Other features of high importance are minimum beam perturbations, and excellent counting stability. These parameters were measured with different neutron converters on different beamlines at the ILL.

### 3.1 Experimental setup

Most of the characterization measurements were done on the CT1 2.5-Å-monochromatic beamline at the ILL at a flux around $10^5$ n·cm$^{-2}$·s$^{-1}$ (Figure 3). The beam monitor to be tested, filled with various gas mixtures, is mounted on a dual-axis translation table and a high efficiency $^3$He detector (96 %) – referred further as reference detector – acts as beam stop downstream and monitor the beam intensity. Unless specified otherwise, the monitor to be characterized is connected to a single preamplifier and a MCA (Multi-Channel Analyser). Motorized neutron absorbing slits located upstream allow to adjust the size of the beam.

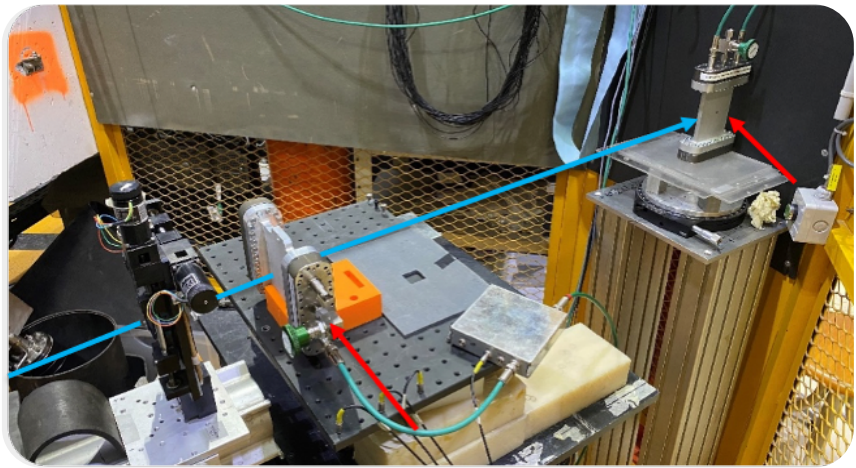

**Figure 3.** Experimental setup to measure uniformity of response and transparency of the monitor. The blue arrow shows the path of the neutron beam. The red arrows locate the reference detector (right) and the scanned monitor (left).

## 3.2 Efficiency and Pulse height spectrum

The gamma-ray signals can degrade the precision of the neutron flux measurement in two ways: the signal amplitude may be above the discrimination threshold and be detected as a false neutron, and, secondly, the gamma-ray interaction rate may be sufficient to induce current leakage between anode and cathodes, resulting in an unstable signal base-line, and unstable trigger efficiency. To be able to measure precisely the neutron flux in a high-gamma background field, the gamma interaction rate must not exceed the limit of stability of the electronics, and the number of gamma-rays detected must be relatively small compared to the number of neutrons detected.

It is also important that the neutron counting rate be sufficiently high to minimize statistical uncertainty, whilst keeping the deviation from counting linearity at an acceptable level (the limit is generally set at 10 %), for all the conditions encountered on the instrument. The mass of neutron converter inside the monitor is specified in such a way that this counting rate limit is reached for the most intense neutron beam used. The uncertainty of the converter mass inside the monitor has to be considered, as well as the uncertainty of the maximum neutron flux of the instrument, in particular for new instruments.

Figure 4 shows a pulse height spectrum measured with the monitor filled with 10 mbar of $^3$He, and 500 mbar of Ar-CO$_2$ (90:10). The peak at channel 1200 illustrates the so-called wall effect: for a majority of events, the decay products, proton and tritium, are lost in the walls before transferring all their kinetic energy to the gas; only a fraction of the interaction energy is detected. Nevertheless, this peak is well separated from the background due to gamma-rays and electronic noise. A stable counting rate can be achieved by setting the signal discrimination threshold between background events and neutron peak (red dashed line on Figure 4).

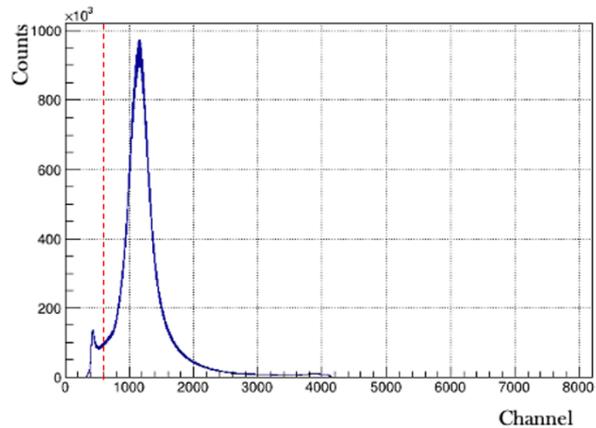

**Figure 4.** Pulse height spectrum of multitube monitors filled with Ar-CO2 and $^3$He. Trigger threshold is represented by the dashed line.

The pulse height spectrum measured with the monitor filled with 400 mbar of N$_2$ and 600 mbar of Ar-CO$_2$ (90:10), is shown in Figure 5. Masking temporarily the monitor with a B$_4$C sheet, we determined the threshold for gamma-ray signals at channel 600 where we set the discrimination threshold of the MCA for the acquisition of the pulse height spectrum. For N$_2$, given the difference of mass between the two emitted particles, most of the reaction energy (93 %) is carried by the proton, and the remaining 7 % by the $^{14}$C. For the neutrons interacting close to the surface of a tube wall and resulting in a $^{14}$C emission toward the tube centre, the energy deposited in the gas is less than the $^{14}$C kinetic energy, which is only 42 keV. The important wall-effect resulting from this type of events corresponds to the large decreasing edge on Figure 5 between channel 600 and 2500. The peak at channel 3000 is equivalent to the one observed for $^3$He; it corresponds to neutron events where the proton transfers part of its energy to the gas and loses the rest in the tube walls. The peaks close to channel 5600 and 5800 are the full absorption peaks, which means that the proton or both the proton and the $^{14}$C were emitted along the tube axis and transferred all their energy to the gas.

– 6 –

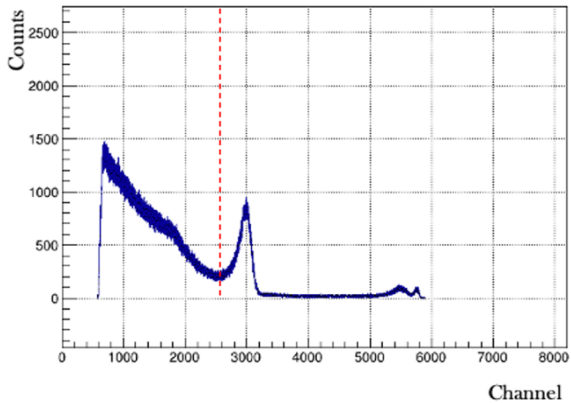

**Figure 5.** Pulse height spectrum of multitube monitors filled with Ar-CO2 and $N_2$. Trigger threshold is represented by the red dashed line.

The signal discrimination threshold should be set so as to minimize the sensitivity of the trigger to variations of HV, amplification gain, or electronics noise. From Figure 5 one can see that the trigger level for optimal stability corresponds to channel 2600 in the PH spectrum. Under these conditions trigger efficiency will be around 20 %, leading to a reduction in the counting dynamic.

As the energy gap between neutrons and gamma-rays is more favourable for $^3$He than for $N_2$, beam monitors filled with $^3$He are intrinsically superior to those filled with $N_2$. The main reason for using $N_2$ is to benefit from its low cross section, which allows for a better definition of the partial pressure of the converter in the final gas mixture. Given the ratio between $^3$He and N capture cross sections, 145 mbar of $N_2$ or 0.1 mbar of $^3$He both correspond to the same value of detection efficiency, $10^{-5}$, which is a typical efficiency value for beam monitoring. To achieve such a low concentration of $^3$He in the detection gas with sufficient accuracy, the $^3$He is diluted over several steps in another gas.

### 3.3 Uniformity of detection efficiency

The response uniformity of a beam monitor is not a decisive quality criterion, as long as the beam remains stable both in shape and energy distribution, and if the only parameter to correct is the overall flux variation. Normalization might be difficult for experiments consisting of different sets of data corresponding to different beam conditions, if they depend on the counting rate measured by a non-uniform beam monitor. For example, if a neutron attenuator consisting of holes in an absorber is used, then, depending on the position of the holes relative to the beam monitor, the beam attenuation measured might be different from the attenuation factor calculated. This can be an issue if data measured with different collimations are to be calibrated according to the theoretical reduction factor. Another situation is when the shape of the beam slowly changes over time due to small variations in the alignment of the elements composing the neutron guide; this may result in variations in the rate measured by the beam monitor which do not reflect beam flux variations.

To evaluate the uniformity of the Multitube monitor, we used the setup described on Figure 3 and collimated the beam with a 3 mm × 3 mm aperture. The sensitive area of the monitor is scanned by moving the translation table in both directions with a pitch of 3 mm. For each position, the counting rate of the monitor was measured, as well as the intensity of the transmitted beam, using the reference detector located 1 m downstream of the beam monitor (Figure 3). The solid angle covered by this reference detector from the Multitube monitor centre is 0.1 % × 4π; although we know that neutrons are not isotropically scattered, we can consider that the rate of scattered neutrons counted by the reference detector is negligible. For each point of the scan, the count rate of the monitor is divided by the one of the reference detector. Then it is multiplied by the known efficiency of the reference detector, which gives a local detection efficiency value for the monitor.

The average measured detection efficiency $2.94\times10^{-3} \pm 0.7$ %. Considering that the detector was filled with 550 mbar of Ar-$CO_2$-$^3$He (86.4;9.6;4) i.e. around 22 mbar of $^3$He, this result is consistent with the expected efficiency ($2.96 \times 10^{-3}$). The measured mean neutron transparency is $97.6 \pm 0.4$ % (Figure 6). This number matches well with the calculated value of 97.7 % (0.3 % of the neutrons absorbed in the gas, and 2 % scattered or absorbed in the monitor material). For both measurements, most of the uncertainty comes from the statistics of the beam positions relative to the inner walls. The attenuation factor of the Multitube beam monitor is a factor of 2 better than the one of other beam monitors of same detection efficiency used at the ILL and commercially available like fission chambers and proportional counters (See reference [1]).

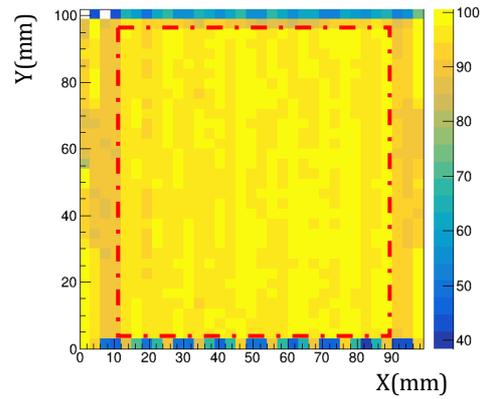

**Figure 6.** 2D plot of Multitube monitor transparency measurement measured in percentage (color bar) of the transmitted beam.

### 3.4 Counting stability

A Multitube monitor has been in continuous operation on the IN16B backscattering spectrometer at the ILL for almost 10 years and has not required any maintenance so far.

On the CT1 beamline at the ILL, two uranium fission chambers (with ~$2\times10^{-3}$ and ~$5\times10^{-4}$ detection efficiency) and one Multitube monitor (with a detection efficiency of $1.4\times10^{-3}$) were stacked together and placed across the neutron beam. The fission chamber with an efficiency of $2\times10^{-3}$ was used as the reference monitor to correct for beam flux variations. The counting rate of the two other beam monitors were divided by the counting rate of the reference monitor, and normalized to the counting rate value measured at the beginning of the experiment.

Figure 7 shows the resulting curve measured over 2 days. We see that the normalization corrects the beam flux variations well, which can be $\pm0.7$ % of the raw mean counting rate over the same period. The fact that the two curves remain flat during a long period is a strong indication that the 3 beam monitors are stable. The curve fluctuations are dominated by statistical uncertainties. They are higher (0.6 %) for the low efficiency fission chamber than for the Multitube (0.2 %), due the higher counting rate of the latter.

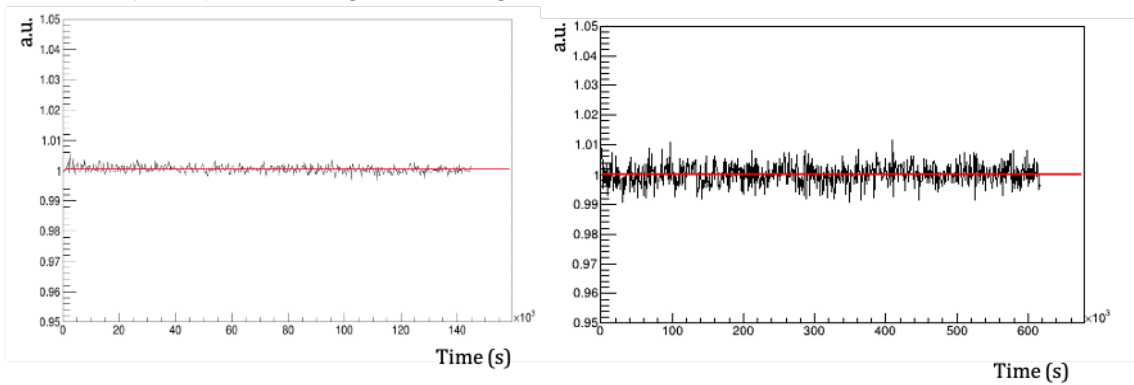

**Figure 7.** Normalization of the monitor rates (multitube on the left and second Uranium on the right) to the first Uranium monitor rates. The red line is the result of the linear fit.

## 3.5 Measurement with low gas pressure

Measurements were performed at the ILL on the H523 neutron guide with a flux of ~$2\times10^7$ n·cm$^{-2}$·s$^{-1}$ distributed over 2.1 cm$^2$. This beam line is 200 times more intense than CT1. The Multitube monitor was located 4 m away from the end of the guide and was irradiated once with the full neutron beam and once with a 20 mm-thick neutron shielding B$_4$C plate in contact with the beam monitor. Neutrons captured by the B$_4$C plate generate gamma-rays that add to the gamma-ray radiation background of the instrument. In order to study the influence of the monitor gas on the gamma-ray sensitivity of the monitor, the monitor was filled with 20 mbar of $^3$He and either 500 mbar or 2 bar of Ar-CO$_2$.

Figure 8 shows that few gamma-ray events are counted in the neutron region (between channels 10000 and 60000) at 2 bar of Ar-CO$_2$, while this is not the case at 0.5 bar. The gamma-ray flux emitted by the B$_4$C plate from neutron capture has been conservatively evaluated to be at least[1] $10^9$ ɤ·s$^{-1}$, assuming that one third of the gamma rays reach the monitor; this gives a maximum sensitivity to gamma at 2 bar close to $10^{-8}$. So far, at 500 mbar, we have not been able to obtain a value since there is no "gamma-ray event" in the region of interest when the detection threshold is set at channel 10000.

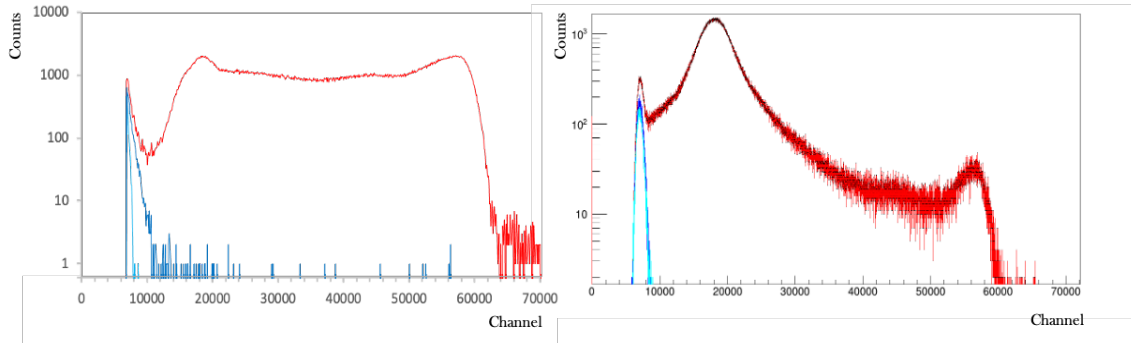

**Figure 8.** Pulse Height Spectra (log. scale) measured for 300 seconds with the beam monitor filled with 2 bar of Ar-CO$_2$ and 20 mbar of $^3$He (left), and 0.5 bar of Ar-CO$_2$ and 20 mbar of $^3$He (right) with neutron beam on (red), off (light blue), and stopped with B$_4$C in front of the detector (dark blue).

Gamma-rays may interact with either the monitor vessel or the detection gas inside it. The gas actually accounts for less than a thousandth of the gamma interactions [5]. Tuning the gas mixture does not help to reduce the gamma interaction rate [6], but it helps to minimize energy deposited in the gas by the resulting photo-electrons or Compton electrons. The charge density of ionisation tracks is lower for electrons generated by gamma-ray events than for heavy particles produced in neutron events; thus, the energy gap between the two types of events can be maximised by a low quenching gas pressure. This is true only for pressure values down to a certain limit: below that, there is no benefit to be gained from reducing the pressure further in terms of gamma sensitivity. We can just conclude that 0.5 bar is more favourable than 2 bar for gamma discrimination.

## 3.6 Test in high gamma rate field

In order to test the beam monitor in an intense gamma field, the pulse height spectrum was measured on instrument PF1b at the ILL. PF1b, instrument mostly dedicated to fundamental physics experiments requiring high statistics, has one of the most intense white cold neutron beam at ILL with a neutron flux of $2\times10^{10}$ n·cm$^{-2}$·s$^{-1}$ at the guide exit with a mean wavelength of 4 Å. As a consequence, the gamma rate in the casemate where the cold neutron guide ends is also one of the highest.

---
[1] Although this value could be much higher, since more gamma rays could come from the guide itself.

The detector was filled with 330 mbar of $CO_2$ and 0.4 mbar of $^3$He, corresponding to a neutron detection efficiency of $8\times10^{-5}$ at 4 Å. Due to neutron absorbing components between the end of the neutron guide and the detector, the average counting rate measured was 250 kHz. The flux of gamma-rays was measured at a position close to the monitor with a gamma-meter and evaluated to be at least $10^{10}$ ɣ·s$^{-1}$. Along the 3 days of measurements, the pulse height spectrum remained very stable with no detectable change in shape, except a variation of ± 0.3 % in the number of counts, which is consistent with the value expected due to variations in reactor power. This result demonstrates the excellent stability of the Multitube beam monitor even in a high gamma field.

The shape of the pulse height spectra in Figure 9 shows excellent energy separation between gamma rays (first peak) and neutrons, suggesting that the monitor would still be operational on a more intense neutron beam.

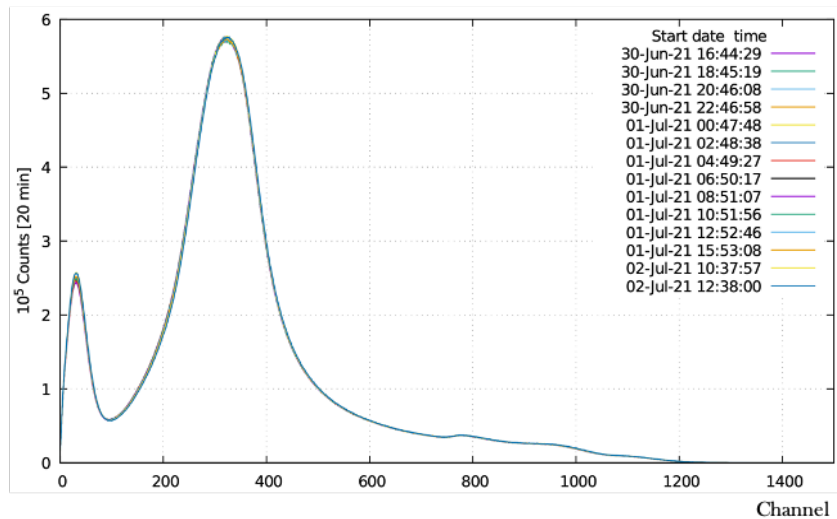

**Figure 9.** Pulse Height Spectrum measured at different times with the Multitube monitor installed on PF1b at ILL.

### 3.7 Counting rate capability and dead time measurement

A high counting rate capability allows for a broad range of neutron beam intensities. This is particularly useful for instruments like IN20 at the ILL, which uses several monochromators providing different neutron flux outputs, or for instruments with intense pulsed beams. A dead-time measurement was performed on IN20 instrument (Figure 10) using two Multitube monitors with a forty-five-fold difference in efficiency between the two.

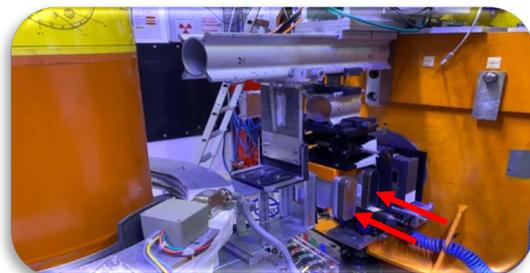

**Figure 10.** Dead time measurement setup on IN20 at ILL using two monitors with different efficiencies

These monitors were irradiated across a 4×4 cm² area (covering 6 tubes) at a flux of $4\times10^7$ n·cm$^{-2}$·s$^{-1}$. From these measurements (Figure 11), performed with different beam collimations, we see that the monitor with high-efficiency loses 10 % of counts at 550 kHz. A dead-time of 230 ns was estimated using the standard non-paralyzable model to fit the curve. This value is consistent with the signal shape delivered by the bipolar preamplifiers used. This dead-time is comparable with the dead time of fission chambers, and a factor of 20 better than standard proportional counter monitors thanks to the low pressure (330 mbar of the gas mix for this experiment) and to the geometry of the Multitube that enhances the collection of charges due to the proximity of the walls.

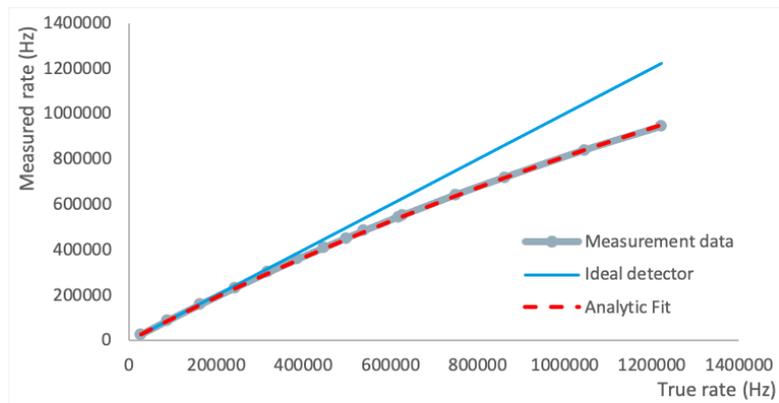

**Figure 11.** Rate measured by the multitube monitor versus the expected true rate of interactions

Another solution to expand the counting range by a factor of 10 is to connect the central tube to an individual read-out channel and the other tubes to a second read-out channel. In the low-intensity mode, the beam intensity is measured by summing the signals of all the tubes, whereas only the central tube is read out in the high intensity mode. This can be implemented easily at almost no extra cost: only one additional HV feedthrough is required.

## 4. B$_4$C film neutron converters

The B$_4$C thin films with natural isotope abundance ($^{nat}$B4C) on 125 μm Al foils (>99.0 wt.% purity) were produced by an industrial deposition system (CC800/9, CemeCon AG) at ESS Detector Coatings Workshop in Linköping, Sweden. Details about the deposition unit and process development for neutron detector coatings of other types have been published elsewhere [7][8][9]. Four $^{nat}$B$_4$C targets (>99.5 wt.% purity) were co-sputtered from cathodes for direct current magnetron sputtering, where the deposition time was varied to achieve different film thicknesses ranging from 25 nm to 400 nm, as listed in Table 1. The depositions were made on 150 mm × 350 mm foils; they were thereafter shaped into strips of 8 mm × 120 mm to be mounted inside the tubes (Figure 12).

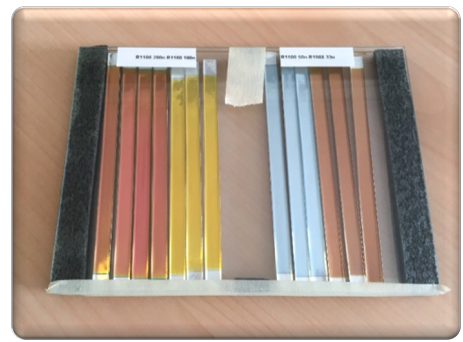

**Figure 12.** B$_4$C-coated aluminum blades that fit the tube geometry of the Multitube monitor

We did the same measurements as the ones described in Section 3.3; we scanned the monitor with the collimated beam all over its surface and evaluated the efficiency of the monitor for each point. Figure 13 shows the distribution of positions measured by charge division, as described in section 2.3, for 10 points of the scan taken at the center of each tube of the monitor during the same acquisition time. In this plot, each peak corresponds to one tube. The difference of intensity between the peaks corresponds to the difference of detection efficiency between the converters inserted in the tubes. The counting rate in each tube is obtained by summing the counts in each peak, divided by the acquisition time. The detection efficiency of each film is thus given by this number divided by the intensity of the neutron beam measured with the reference detector. The converter thickness corresponding to this measured efficiency was calculated and we compared the result to the specified converter thickness. Table 1 shows a good agreement between the two. The variation of the efficiency along the length of the converter was also measured using successive measurement data points along each tube axis: the maximal deviation from the mean value is 3 %. The fact that the two tubes with no $B_4C$ film are counting neutrons is attributed to residual $^3$He contaminant in the gas filling system. The detection efficiency of $2.5 \times 10^{-6}$ measured with these tubes corresponds to a $^3$He partial pressure of $10^{-2}$ mbar. The measurement was repeated with a different Multitube without $B_4C$ samples, and after pumping the gas filling system during several hours. No neutrons were detected during an acquisition time of 10 minutes.

**Table 1.** Measured efficiencies of the Multitube monitor equipped with Boron-coated blades

| Thickness (nm) | 400 | 200 | 100 | 50 | 25 | 0 | 200 | 100 | 50 | 0 |
|---|---|---|---|---|---|---|---|---|---|---|
| Detection efficiency (×10$^{-3}$) | 4.39 | 2.2 | 1.13 | 0.55 | 0.29 | 0.003 | 2.16 | 1.11 | 0.556 | 0.007 |
| Calculated Thickness (nm) | 393.5 | 196.5 | 101.1 | 48.6 | 25.9 | 0.2 | 193.3 | 99.4 | 49.7 | 0.6 |

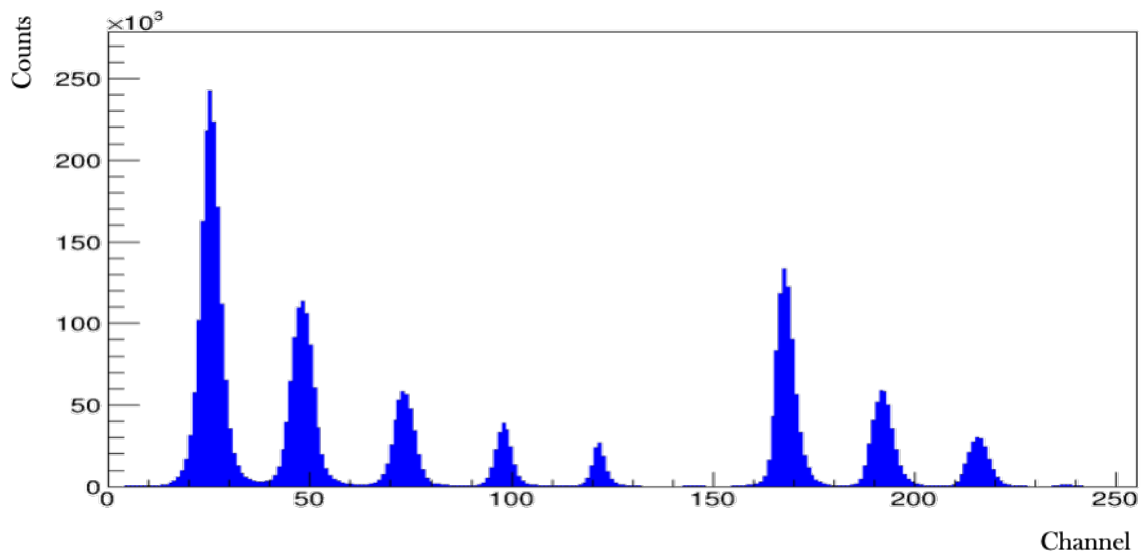

**Figure 13:** Distribution of counts measured with the 1-D configuration. Each peak corresponds to a tube containing an aluminium blade coated with a B4C film of a given thickness. The integral of each peak was used to calculate the values in Table 1.

We also made $^{nat}B_4C$ films of 5 nm and 10 nm thickness; For half of those we added during the deposition an electroformed micro-patterned mask with 27 % aperture. The latter films have then a regular pattern with alternatively $B_4C$ coating and absence of coating resulting in an average effective thickness of ~1.25 and ~2.5 nm. Preliminary results with these films show that we are able to reach efficiencies down to $1\times10^{-5}$ with this method.

## 5. Beam imaging

As discussed in Section 2.1.3, the Multitube monitor can also be configured to localize each neutron in two-dimensions. As a proof of concept, the monitor was irradiated with the neutron beam on CT1 both with a slit mask and without a mask. The detector was filled with 1 bar of $^3$He and 0.5 bar of Ar-$CO_2$ (90:10), and the amplification gain was increased by a factor of around 10 to reduce the contribution of electronics noise in the spatial resolution.

A calibration is required to convert the raw data into positions. The data used for calibration are recorded by moving the Multitube monitor on the CT1 translation table, in front of the collimated beam, so that each tube is irradiated at different positions along its length.

Figure 14 shows the images obtained after calibration on CT1 without and with a mask made of several slits equally spaced at a distance of 20 mm.

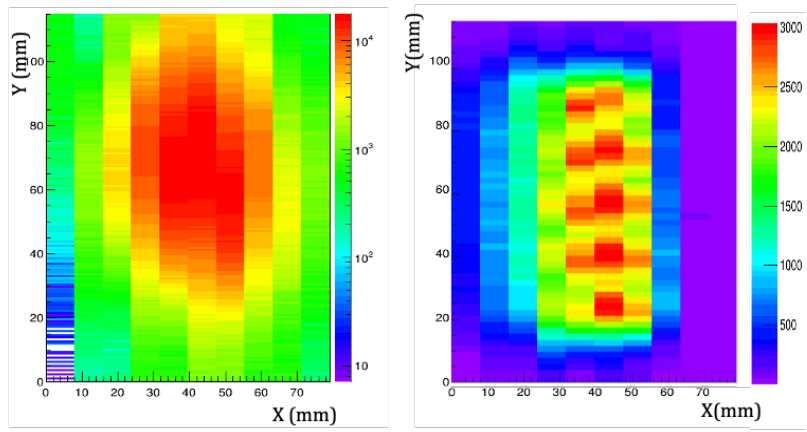

**Figure 14:** 2-D images obtained with the Multitube on CT1. The image on the left shows the image of the full beam at the exit of the beamline without collimation. The image on the right shows the same but with a neutron absorber mask made of slits with a pitch of 20 mm (The two last tubes are masked by the support of the mask).

The resolution along the tubes is limited by the low stopping power of the gas, which could be improved. The resolution was evaluated to 14 mm, thanks to another mask made of 5 slits equally spaced by 14 mm perpendicular to a single tube of the monitor. The masked monitor was irradiated once with the full beam and successively with only one slit allowing neutrons to go through (the remaining ones were masked with an additional layer of $B_4C$). On Figure 15, we can see the positions measured with only slit are well separated at Full Width at Half-Maximum.

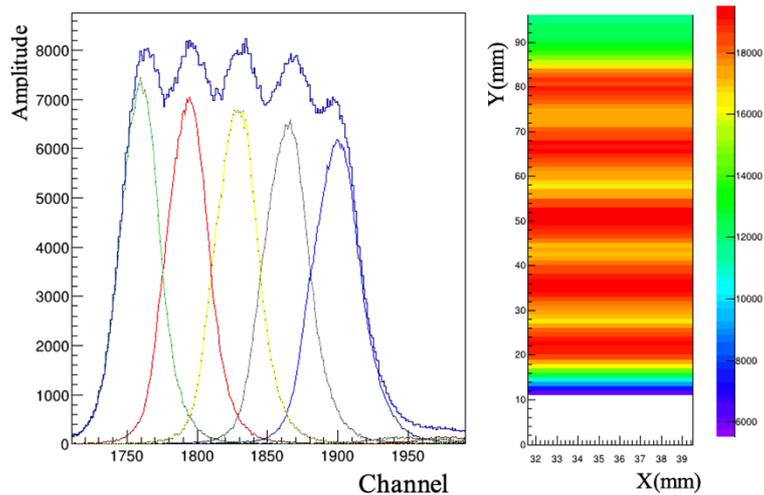

**Figure 15.** Measured positions of neutron events along one tube (left) and corresponding "2-D" image (right). The top blue curve on the left plot is obtained with the full beam on the monitor equipped with the 5 slits mask spaced equally by 14 mm distance. The colored curves (green, red, yellow, grey, light blue) are obtained for only one slit allowing neutrons to come through.

To improve the resolution in the other direction, tubes with smaller section can be used; a Multitube with 28 tubes of 2.6 mm × 4 mm internal section and 105 mm long spaced by a 3 mm pitch has been manufactured (**Figure 16**), to study the potential of this approach. The sensitive area is 75 mm × 84 mm.

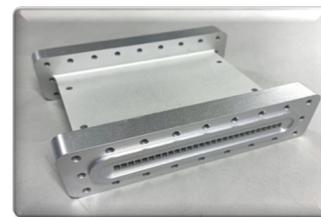

**Figure 16:** Vessel body of the new Multitube with 28 small tubes

## 6. Conclusion

For instruments requiring precise monitoring of the beam and short acquisition times, a high counting rate capability is crucial, whilst avoiding saturation. This is particularly important for instruments using pulsed neutron beams -- those on spallation neutron sources or equipped with choppers for example. The possibility offered by the Multitube monitor of changing the sensing gas to match the specified efficiency is an advantage.

The Multitube beam monitor can be used in vacuum, has a good magnetic compatibility, and can withstand very high incident neutron beam intensity of at least $4\times10^{11}$ n·cm$^{-2}$·s$^{-1}$, at a counting rate of 550 kHz with 10 % loss. We have shown that the beam perturbation is lower than compared to other beam monitors used in neutron research facilities, namely fission chambers and proportional counters. The counting stability of the Multitube is excellent, as well as its response uniformity, both for detection efficiency, and for beam transparency. The monitor also provides 1-D and 2-D read-out features useful for the alignment of the instrument, and to control the uniformity of the beam. The use of low pressure and optimised geometrical set-ups also improves gamma-ray discrimination. Several Multitube monitors have been fabricated for the ILL instruments ( SHARP, D16, D20, PF1B, and SuperSun), and several others are planned for the near future.

**Acknowledgments**

This work was carried out at the ILL, in a collaborative effort to identify the best options for neutron beam monitoring at the ESS and ILL facilities.